\documentclass[prd,aps,preprintnumbers,nofootinbib]{revtex4}

\def\beqra{\begin{eqnarray}}
\def\eeqra{\end{eqnarray}}
\def\beq{\begin{equation}}
\def\eeq{\end{equation}}

\def\ab{\bar{a}}
\def\rhb{\bar{\rho}}
\def\Phib{\bar{\Phi}}
\def\Psib{\bar{\Psi}}

\def\vp{\varphi}

\def\gmn{g_{\mu\nu}}
\def\tgmn{ \tilde{g}_{\mu\nu}}

\def\agt{\stackrel{>}{\sim}}
\def\alt{\stackrel{<}{\sim}}
\begin{document}

\preprint{DESY 06 -- 076} 

\title{Einstein and Jordan frames reconciled: a frame-invariant approach to scalar-tensor cosmology}
\author{Riccardo Catena}
\address{Deutsches Elektronen-Syncrotron DESY, 22603 Hamburg, Germany}

\author{Massimo Pietroni}
\address{INFN, Sezione di Padova, via Marzolo 8, I-35131, Padova, Italy}

\author{Luca Scarabello}
\address{Dipartimento di Fisica Universit\`a di Padova and INFN, Sezione di Padova, via Marzolo 8, I-35131, Padova, Italy}

\begin{abstract}
Scalar-Tensor theories of gravity can be formulated in different frames, most notably, the Einstein and the Jordan one. While some debate still persists in the literature on the physical status of the different frames, a frame transformation in Scalar-Tensor theories amounts to a local redefinition of the metric, and then should not affect physical results. We analyze the issue in a cosmological context. In particular, we define all the relevant observables (redshift, distances, cross-sections, ...) in terms of frame-independent quantities. Then, we give a frame-independent formulation of the Boltzmann equation, and outline its use in relevant examples such as particle freeze-out and the evolution of the CMB photon distribution function. Finally, we derive the gravitational equations for the frame-independent quantities at first order in perturbation theory. From a practical point of view, the present approach allows the simultaneous implementation of the good aspects of the two frames in a clear and straightforward way.

\end{abstract}

\maketitle

\maketitle
\section{Introduction}
The evidence for Dark Energy has revived the interest in modifications to General Relativity (GR). Among these theories, Scalar-Tensor gravity (ST) \cite{bd} represents a good benchmark 
to accommodate new ultra-light degrees of freedom possibly responsible for the accelerated expansion of the universe. 

From a phenomenological point of view, it respects local Lorentz invariance and the universality of free fall of test bodies. Moreover, the post-Newtonian parameters $\gamma-1$ and $\beta-1$, parameterizing the deviations from GR, are expressed in terms of a single function, thus allowing a straightforward confrontation of the theory with solar system tests of gravity \cite{pn,dam,cassini}. 

From a theoretical point of view, this class of theories is large enough to accommodate a vast range of possible extensions of GR in which new scalar fields are present in the gravitational sector; from extra-dimensional radions and string theory moduli, to $f(R)$ theories of gravity. Moreover, in a ST context, ultralight scalar fields are technically natural. Indeed, general covariance implies that the contribution of radiative corrections from the (visible and dark) matter sector to the scalar field mass is at most of order $\Lambda^4/M_p^2$, $\Lambda$ being the cosmological constant. Thus, the lightness of the scalar field is just a manifestation of the smallness of the cosmological constant, or of the curvature of the universe \cite{choi}. 

Finally, in a cosmological setting, it has been pointed out that an intriguing attraction mechanism towards GR \cite{GRattr} could be operative under very generic conditions, including the case of a runaway potential suitable for DE \cite{Bartolo,Catena,Esposito}. Therefore, these theories may differ considerably from GR at high redshifts and at the same time fulfill the stringent bounds coming from solar system tests \cite{cassini} today.

ST theories can be formulated in different guises. In the so-called `Jordan frame', the Einstein-Hilbert action of GR is modified by the introduction of a scalar field\footnote{ST theory can be generalized with the introduction of many scalar fields \cite{bd}. In order not to overload the notation, in this paper we will consider a single field, but our results are easily generalizable to the multi-field case.} with a non-canonical kinetic term and a potential. This field replaces the Planck mass, which becomes a dynamical quantity. On the other hand, the matter part of the action is just the standard one.

By Weyl-rescaling the metric, one can express the ST action in the so called `Einstein Frame'. In these new variables, the gravitational action is just the Einstein-Hilbert one plus a scalar field with canonically normalized kinetic terms and an effective potential. On the other hand, in the matter action the scalar field appears, through the rescaling factor multiplying the metric tensor everywhere. As a consequence, the matter energy-momentum tensor is not covariantly conserved, and particle physics parameters, like masses and dimensionful coupling constants appearing in the lagrangian are space-time dependent.

It is a general fact that physics is invariant under a local redefinition of field variables -- in this case, the Weyl-rescaled metric.
 Nevertheless, this invariance is not fully exploited in the literature, where some confusion also exists about the physical status of the different frames. Most authors prefer to work in the Jordan frame, which is also referred to as the `physical' one. The advantage of this frame is that all the particle physics' properties, {\it i.e} masses, coupling constants, decay rates, cross sections, etc. can be computed straightforwardly, since the matter action is just the standard one. On the other hand, the gravitational equations are more involved than in GR, since the scalar is non-trivially mixed to the metric tensor. 

Working in the Einstein frame is easier for what concerns the gravitational equations, but the connection with particle physics is not so direct as in the Jordan frame, since, for instance, the electron mass appearing in the lagrangian is space-time dependent. So, many authors use the Einstein frame as a mathematical tool to solve the field equations, and then translate back the results in the Jordan frame to compare with observations. For some recent applications along these lines, see, for instance \cite{Catena,Coc,Schimd,Martin}. Non-linear approaches have also been discussed \cite{Perrotta,Matarrese}. 

While both these procedures are correct, a general discussion of the frame-invariance of physics in the ST theories in a cosmological setting is still missing. In particular, while it is quite straightforward to go back and forth from one frame to the other when the barotropic fluid approximation for matter holds, it is not so clear how to do it when Boltzmann equations have to be employed. For instance, the epoch of decoupling of some interaction in the expanding universe, expressed in conformal time, should be derived independently on the frame. But the standard rule of thumb, that is, 
\beq
\Gamma \alt H/a\,,
\eeq
with $\Gamma$ the reaction rate and $H$ the Hubble parameter, is not a frame-invariant relation. 


The purpose of this paper is to formulate a frame-invariant approach to discuss ST cosmology. Following Dicke \cite{Dicke, barcellona}, we will start by the observation that the physical observables are dimensionless ratios between physical quantities and the appropriate units of measure\footnote{Strictly speaking, this argument applies only to local physical quantities independent of metric derivatives.}. These numbers are frame-invariant, and should therefore be expressible in terms of frame-invariant quantities. We will discuss the main observables in cosmology (redshift, distances, CMB temperature perturbations, \ldots) and particle physics (masses, cross sections, rates, \ldots) and express them in terms of frame-invariant combinations of the theory parameters and variables, and the units. We will discuss metric perturbations and define a frame-invariant phase space and distribution function. This will enable us to write down a frame-invariant Boltzmann equation to discuss processes relevant for cosmology, like particles freeze-out, the Sachs-Wolfe effect, and matter-radiation decoupling. Finally, we will derive the equations of motion for the frame invariant quantities at first order in metric perturbations.

The plan of the paper is as follows. In sect.~\ref{frt} we will introduce frame-transformations.
In sect.~\ref{fipp} we will discuss how frame-invariant results can be obtained for rates, cross sections, and so on. The cosmological background observables, that is redshift and (angular and luminosity) distance will be discussed in sect.~\ref{fifrw}. Then, we will consider scalar perturbations in the Newtonian gauge. We will see that a frame transformation amounts to a change of gauge, and will therefore define frame-invariant scalar metric perturbations. This will enable us to define a frame-invariant phase space and energy-momentum tensor (sect.~\ref{fip}). The Boltzmann equation will be introduced in sect.~\ref{Boltzmann}, where its application to particles' freeze-out and CMB photons will be outlined. 
Finally, in sect.~\ref{EoM}, we will discuss the ST dynamics. We will write down the ST action in terms of frame-invariant quantities only, and write down the corresponding equations of motion at first order. In the appendix, the case of the synchronous gauge and of a generic gauge will be discussed.

\section{Frame transformations}
\label{frt}
In general, a frame transformation is a rescaling of  the metric $\gmn$, of the form
\beq \tgmn = e^{-2 f} \gmn\,,
\label{trans}
\eeq
with $f=f(x^\mu)$ and real, and the space-time coordinates $x^\mu$ ($\mu=0,\ldots,3$) are kept fixed. The diffeomorphism-invariant space-time interval then gets transformed as
\beq
d\tilde{s}^2=\tgmn dx^\mu dx^\nu =  e^{-2 f} \gmn dx^\mu dx^\nu = e^{-2 f} ds^2\,.
\label{ds2}
\eeq
Considering time-like and space-like intervals, we see that proper times, $dt_0=|ds|(\mathrm{tim elike})$, and proper lengths, $dl_0=|ds|(\mathrm{spacelike})$, transform as
\beq
d\tilde{t}_0=e^{-f} dt_0\,\;\;\;\;\;\;\;d\tilde{l}_0=e^{-f} dl_0\,.
\label{timelength}
\eeq
The above transformations can be seen as the relations between the clocks and rods in two different systems of units \cite{Dicke, Nuovecits}. Notice that the above transformation laws are, in general, {\em space-time dependent, since so is the function $f(x^\mu)$}. Such transformations are not common in GR and in standard physics in general, where they are of no practical use. However, in ST the situation changes dramatically. For instance, one could consider two different units of length, say,  the size of an atom and the radius of a small Schwarzschild black hole. The ratio between these two physical lengths, which is constant in GR, is in general space-time dependent in ST. Therefore, in this kind of theories,  the operational definition of units implies local transformations. 

Since $ds^2=0$ is a frame-invariant condition, the speed of light is also invariant under the transformation (\ref{timelength}). Moreover, we will {\em impose} that the Planck constant $\hbar$ is also invariant. This amounts to transforming units of mass (and energy) as 
\beq
\tilde{M}=e^f \,M\,,
\eeq
and in leaving actions invariant (but, in general  not covariant!).
It should be emphasized that this is by no means a unique choice. One could for instance consider frame transformations in which 
$\hbar$ varies and the Newton constant is kept fixed. However, our prescription is the one realized in ST theories, which are the main focus of this work. 

Since -- in the particular class of local transformations we are considering -- units of time, length, and inverse masses transform in the same way, we will consider a single dimensionful unit, length. The generic unit length will be indicated by $l_R$. It transforms according to the space-time dependent relation of eq.~(\ref{timelength})
\beq
\tilde{l}_R= e^{-f} l_R\,.
\eeq
Then, a generic {\em local} physical quantity, $Q$, having the dimensions [Mass]$^a$ [Length]$^b$, and [Time]$^c$, will transform according to
\beq
\tilde{Q} = e^{f(a-b-c)}\,Q\,.
\eeq
The scaling above holds as long as the quantity $Q$ does not depend on derivatives of the metric. The choice of the unit length is dictated, as usual, by practical convenience. For instance, one could use a reference atomic wavelength, the inverse physical mass of some particle, or the Planck length.

Physics does not depend on which particular clock or rod one adopts. In the following, we will discuss this independence in the framework of Friedmann-Robertson-Walker cosmology, but of course this is a  general property \cite{Dicke}.

\section{Frame-invariant particle physics}
\label{fipp}
The particle physics action (the underlying theory being the Standard Model, or any of its extensions) has the form
\beq
\int d^4x\,\sqrt{-g}\, {\cal L} = \int d^4x\,\frac{\sqrt{-g}}{l_R^4}\, l_R^4\, {\cal L}\,.
\eeq

Since both the action and the combination $\sqrt{-g} / l_R^4$  are frame-invariant,   so is  the product $l_R^4 \,{\cal L}$. It is convenient to construct a lagrangian $\tilde{\cal L}$ such that  
\beq 
\tilde{l}_R^4 \, \tilde{{\cal L}}=l_R^4 \,{\cal L}+\cdots\,,
\label{puntini}
\eeq where the dots represents terms containing space-time derivatives of $l_R$ and $\tilde{l}_R$. This is achieved transforming the parameters and fields in ${\cal L}$ as follows:
\beqra
&& l_R^n \,\lambda_n = \tilde{l}_R^n \, \tilde{\lambda}_n\,,\;\;\; l_R \,\phi = \tilde{l}_R \,\tilde{\phi}\,,\;\;\;  l_R^{3/2} \,\psi = \tilde{l}_R^{3/2} \,\tilde{\psi}\,,\nonumber\\
&& A_\mu = \tilde{A}_\mu\,,\,\;\;\; l_R\, \gamma^{\mu} = \tilde{l}_R\, \tilde{\gamma}^\mu\,,
\label{fi}
\eeqra
where $\lambda_n$ is a generic coupling of canonical dimension $n$ (e.g. $m^2\, \phi^2 \equiv \lambda_2\, \phi^2$, $h\, \phi^4 \equiv \lambda_4 \,\phi^4$, and so on), $\phi$, $\psi$, and $A_\mu$ are scalar, spinor and vector fields, respectively, and $\gamma^\mu$ are the Dirac gamma matrices. 

In particular, the mass parameters appearing in the lagrangian  can be constant in one frame and space-time dependent in all the other ones, as $\tilde{m} = l_R/\tilde{l}_R\, m = e^f \, m$. In ST gravity masses and couplings are constant in the Jordan frame and space-time dependent in the Einstein one. 

In general, the space and time scales of variation of the function $f=\log(l_R/\tilde{l}_R)$ depend on the model and can be determined by solving the full set of equations of motion. We will make the assumption that these scales are of cosmological --or at least astrophysical -- size, and in any case much larger than particle physics interaction times and effective ranges \cite{Perrotta,Matarrese}. So, in computing transition amplitudes, decay rates and cross sections, the particle physics parameters adiabatically adjust their relations in eq.~(\ref{fi}) to the local values of the functions $f$, $l_R$ and $\tilde{l}_R$, up to corrections of $O(\lambda_{PP}/L)$, $\lambda_{PP}$ being a typical particle physics interaction range and $L$ the typical scale of variation of $f$. Then, one can compute all the relevant observables in a frame-independent way, following the usual rules of quantum field theory and using the frame-invariant combinations of eq.~(\ref{fi}). The results are frame-independent decay rates, cross sections, etc., given by
\beq 
l_R\,\Gamma = \tilde{l}_R\,\tilde{\Gamma}\,,\;\;\;\;\;\;l_R^{-2} \,\sigma =\tilde{l}_R^{-2} \,\tilde{\sigma}
\,,\;\;\;\;\;\; \cdots\,.
\label{ncs}
\eeq
The above quantities are the true observables, that is, in any frame, the dimensionless combinations between the $\Gamma$'s, $\sigma$'s, ..., and the appropriate powers of the standard rod length.

\section{Frame-invariant FRW cosmology: background observables}
\label{fifrw}
We will consider the background FRW metric
\beq 
ds^2= -a^2(\tau) (d\tau^2- \delta_{ij}dx^i dx^j)\,,
\label{FRW}
\eeq
where $\tau=x^0$ is the conformal time, $\delta_{ij}$ is the delta-function, and latin indices run from 1 to 3. We will assume that the function $f$ defining the frame transformation (\ref{trans}) can be expanded as
\beq
f(\tau, x^i) = \bar{f}(\tau) + \delta f(\tau, x^i)\,,
\label{deltaf}
\eeq
where $ \delta f$ can be treated as a perturbation of the same order as the metric perturbations. Then, the scale factor in the other frame is given by
\beq
\tilde{a}(\tau)=e^{-\bar{f}(\tau)} a(\tau)\,.
\label{scalefactors}
\eeq
\subsection{Redshift and temperature}
One of the basic cosmological observables is the redshift of photon wavelengths. Using the metric (\ref{FRW}) one gets the standard result that a photon traveling through the cosmos, which, at time $\tau_i$ had wavelength $\lambda(\tau_i)$, at a later time $\tau_f$ would have a wavelength
\beq
\lambda(\tau_f)=\lambda(\tau_i) \frac{a(\tau_f)}{a(\tau_i)}\,.
\eeq
Looking at the transformation (\ref{scalefactors}) we see that the ratio $\lambda(\tau_f)/\lambda(\tau_i)$, which is usually defined as the cosmological redshift, is not a frame-invariant quantity. This should be no surprise, since this ratio is not what is actually measured. Instead, the physical quantity is the dimensionless ratio between the wavelength of the -- emitted or absorbed-- photon and some reference length, measured in the laboratory.  Then, the frame-invariant redshift can be defined using a frame-invariant combination such as
\beq
\frac{\lambda(\tau_0)}{\bar{l}_R(\tau_0)} \frac{\bar{l}_R(\tau)}{\lambda(\tau)} = \frac{a(\tau_0)}{a(\tau)}\frac{\bar{l}_R(\tau)}{\bar{l}_R(\tau_0)} \;,
\label{redshift}
\eeq
where the bar denotes the spatial average. 
In order to give an operative definition of redshift, the unit $l_R$ has to be specified. In practice, a reference atomic wavelength is chosen, which we will indicate with $l_R=l_{at}$. In principle, different reference wavelengths could have different space-time dependences, thus leading each to a different definition of redshift. However, in ST theories this is not the case, as they all have constant ratios one another. Therefore, in these theories, the redshift can be defined unambiguously as
\beq
1+z(\tau)  \equiv \frac{a(\tau_0)}{a(\tau)}\frac{\bar{l}_{at}(\tau)}{\bar{l}_{at}(\tau_0)} \;.
\label{redshift_at}
\eeq

The standard relation between the redshift and the scale factor, {\it i.e.} $1+z = \tilde{a}(\tau_0)/\tilde{a}(\tau)$ is recovered only in that frame in which the reference wavelength $\tilde{l}_{at}$ is constant in time and space. In ST theories this is the case of the Jordan frame, whereas, in terms of the scale factor of any other frame one has $1+z = a(\tau_0)/a(\tau) \exp(\bar{f}(\tau) - \bar{f}(\tau_0))$, $f$ being the function connecting the frame under consideration with the Jordan one, according to eq.~(\ref{trans}).
It should be stressed that the reason for the Jordan frame to be singled out from all the possible ones, is a matter of practical utility, namely, the choice of $l_{at}$ in the definition of eq.~(\ref{redshift_at}), but has nothing to do with it having a better physical status than the others. In principle, a physicist could decide to measure the wavelength of cosmological photons in Planck units. In this case, his definition of redshift would have the standard relation to the scale factor of the Einstein frame, not the Jordan one.

Since the comoving coordinate volume is frame-invariant, the total entropy per comoving coordinate volume is a frame-invariant quantity.
In the perfect fluid approximation it is given by the standard expression
\beq
S= \frac{(r a)^3 (\rho+p)}{T}\,,
\eeq
where $r$ is the comoving radius, $\rho$ and $p$ the energy density and the pressure, and $T$ is the temperature. 
Recalling the definition of the energy-momentum tensor for matter,
 \beq
 T_{\mu\nu} =-\frac{2}{\sqrt{-g}} \frac{\delta S_M}{\delta g^{\mu\nu}} \,,
 \eeq
 with $S_M$ the matter action, we get the transformation laws,
 \beq 
 \tilde{T}^\mu_\nu = e^{4f} T^\mu_\nu\,.
 \label{Ttrans}
 \eeq
It should be noted that the above relation is valid as long as the matter action $S_M$ does not depend on derivatives of the metric, as is the case in ST theories.
In the case of barotropic fluids with background energy $\rho=-T^0_0$ and pressure $p\, \delta^i_j = -T^i_j$, we have
 \beq
 \tilde{\rho}= e^{4\bar{f}} \rho\,,\,\,\,\,\,\,\,\tilde{p}= e^{4\bar{f}} p\,.
 \label{perho}
 \eeq
From the frame-invariance of the comoving entropy and using eqs.~(\ref{scalefactors}) and (\ref{perho}), we get that the dimensionless combination $T l_R$ is frame invariant, and that, choosing again $l_R=l_{at}$, the temperature-redshift relation is
\beq
\frac{T(\tau) \bar{l}_{at}(\tau)}{T(\tau_0) \bar{l}_{at}(\tau_0)}=1+z(\tau)\,.
\eeq
Notice that $T\sim 1/a$, in any frame.
\subsection{Distances}
The basic quantity entering the definition of the different distance indicators used in cosmology is the comoving distance traveled by a light ray emitted at redshift $z$, $r(z)$. This is obtained using the (frame-invariant) condition for photon geodesics, {\it i.e.} $ds^2=0$, that is, using eqs.~(\ref{FRW}) and (\ref{redshift_at}),
\beq
 r(z)=\int_0^r dr=\int_a^{a_0} \frac{da}{\dot{a}} = \int_0^z \frac{dz}{1+z}\left(\frac{\dot{a}}{a} - \frac{\dot{\bar{l}}_{at}}{\bar{l}_{at}} \right)^{-1}\,,
 \label{confdist}
\eeq
where the overdot indicates a derivative with respect to the conformal time $\tau$.

The redshift-dependence of the frame-invariant combination $\dot{a}/a - \dot{\bar{l}}_{at}/\bar{l}_{at}$ appearing at the denominator can be computed in any frame. In ST gravity, the $\dot{\bar{l}}_{at}/\bar{l}_{at}$  term vanishes in the Jordan Frame and one needs to solve the Friedmann equations for the scale factor in that frame. On the other hand, in the Einstein frame, one needs also the time-dependence of the field $\bar{f}$ relating the two frames, since $\dot{\bar{l}}_{at}/\bar{l}_{at}=\dot{\bar{f}}$ in this frame. For practical purposes it may be convenient to work in the Einstein frame, since its equations are simpler.

The angular distance of an object of proper diameter D at coordinate r, which emitted light at time $\tau$ (and redshift $z(\tau)$) is given by
\beq
d_A \equiv \frac{D}{\delta }\frac{\bar{l}_R^0}{\bar{l}_R(z)} = a_0\, r(z) (1+z)^{-1}\,,
\eeq
where $\delta$ is the observed angular diameter today, and we have defined $\bar{l}_R^0\equiv \bar{l}_R(\tau_0)$ and $a_0\equiv a(\tau_0)$. In the definition above, we had to keep track of the possibility that the length of the standard rod $l_R$ evolves in time in the frame under consideration. Since $a_0/\bar{l}_R^0$, $z$, and $r(z)$ are all frame invariant, so is also the measure of $d_A$ expressed in terms of the present value of the unit length $\bar{l}_R^0$, that is, the ratio $d_A/\bar{l}_R^0$.

Analogously, the luminosity distance can be defined in a frame-invariant way as
\beq
\frac{d_L^2}{{l_R^0}^2}\equiv\frac{{\cal L}\, \bar{l}_R(z)^2}{4\pi \,{\cal F}\, {l_R^0}^4}\,,
\eeq
where ${\cal L}$ is the luminosity (energy per unit time) of the object at redshift $z$ and ${\cal F}$ is the energy flux measured today.
One can verify that the angular and luminosity distances defined above satisfy the standard relation
\beq 
d_L(z) = (1+z)^2 d_A(z)\,.
\label{dLdA}
\eeq


Finally, the number counts of objects (galaxies, clusters, \ldots) as a function of redshift measure the 
frame-invariant observable
\beq
\frac{d N}{d z} = n_c(z)\, r(z)^2 \left(\frac{\dot{a}}{a} - \frac{\dot{\bar{l}}_{at}}{\bar{l}_{at}}\right)^{-1}  \frac{dz}{1+z}\,d\Omega \, ,
\eeq 
where $n_c(z)$ is the comoving number density.

\section{Frame-invariant perturbations}
\label{fip}
Now we include first order perturbations of the metric and of the function $f$, eq.~(\ref{deltaf}). 
We will work in Newtonian gauge, leaving to the Appendix the extension to the synchronous gauge and to a generic gauge. 

The line element in a generic frame is given by
\beq
ds^2 = a^2(\tau) \left[-(1+2\,\Psi)d\tau^2 + (1-2\,\Phi)\delta_{ij}dx^i dx^j\right]\,,
\label{dsscale}
\eeq
with $\Psi$ and $\Phi$ two scalar functions of space-time. Considering also the fluctuation of $l_R$, $l_R= \bar{l}_R + \delta l_R$, we can write down the frame-invariant line element as
\beqra
&&
ds^2/l_R^2 =  a^2(\tau)/\bar{l}_R^2 \left[-\left(1+2\,\Psi - 2\,\frac{\delta l_R}{l_R}\right)d\tau^2 \right.\nonumber\\
&& \qquad\qquad \qquad\quad\qquad\qquad\left.+ \left(1-2\,\Phi- 2\,\frac{\delta l_R}{l_R}\right)\delta_{ij}dx^i dx^j\right]\,,
\label{dsinv}
\eeqra
which still has the form of an invariant line element in a Newtonian gauge, with frame-invariant scale factor and potentials 
\beq
\bar{a}\equiv a/\bar{l}_R\,,\;\;\;\;\;\bar{\Psi}\equiv \Psi -\frac{\delta l_R}{l_R}\,,\;\;\;\;\;\bar{\Phi}\equiv\Phi +\frac{\delta l_R}{l_R}\,.
\label{fisp}
\eeq
All the physical observables, up to first order, must depend on the above quantities.
In the previous section we have seen already how it works at zeroth order. 
In the following we will extend the program to first-order.

\subsection{Frame invariant geodesics}
It is convenient to work with comoving coordinates, since they are frame-invariant. Then, besides space-time coordinates $x^0=\tau$ and $x^i$, we will also consider the conjugate momenta. For a particle of mass m they are given by
\beq
P_\mu = m \,g_{\mu\nu}\frac{dx^\nu}{ds}\,,
\label{momentum}
\eeq
where $ds\equiv \sqrt{-ds^2}$. 

As we have seen in sect.~\ref{fipp}, the lagrangian mass of a particle is not a frame-invariant quantity, $\tilde{m} = l_R/\tilde{l}_R=e^f\,m$, independent of whether the particle is a scalar, spinor or a vector.

Taking into account the frame dependence of the mass, one can verify that the canonical momenta (\ref{momentum}) with low indices are frame-invariant.

The geodesic equation for a particle with space-time dependent mass is
\beq
P^0 \,\frac{dP^\mu}{d\tau}+ \Gamma^{\mu}_{\lambda\sigma}\,P^{\lambda}\,P^{\sigma}= - m\, \partial_\sigma\,m \,g^{\sigma\mu}\,.
\eeq
Using the metric (\ref{dsscale}) we arrive at the equation for the frame-invariant momentum $P_i$,
\beq
\frac{dP_i}{d\tau} - P_0 \,\partial_i(\Psi+\log\,m) = 0\,,
\label{geod}
\eeq
which is manifestly frame-invariant since $\partial_i(\Psi+\log\,m)=\partial_i(\bar{\Psi}+\log\,\bar{m})$, where $\bar{m}=l_R \,m$.

In some applications, such as the Boltzmann equation entering the computation of the CMB spectra, see sect. ~\ref{Boltzmann},  it is convenient to eliminate the perturbations from the definition of the momenta by going to new variables $q_i$ and $\epsilon$ \cite{MB}, which can be defined in a frame-invariant way as
\beqra
&&P_i=  (1-\bar{\Phi})\, q_i\,,\nonumber\\
&&P_0= - (1+\bar{\Psi})\, \epsilon\,.
\label{qeps}
\eeqra
Writing $q_i=q\, n_i$, with $n_i n_j \delta^{ij}=1$, one can verify the relation $\epsilon=[q^2+\bar{a}^2 \bar{m}^2]^{1/2}$. The geodesic equation in these variables have the standard form
\beq
\dot{q}=q \frac{\partial\;}{\partial\tau}\bar{\Phi} - \epsilon\, n_i \frac{\partial\;\;}{\partial x^i} \bar{\Psi}\,,
\label{geoq}
\eeq
which is valid also in the massless case $\epsilon=q$.

Photon trajectories are given by the frame-invariant equation $ds^2=0$. As a consequence, the expressions for the deflection angles due to weak lensing depend on the frame-invariant combination $\Psi+\Phi=\bar{\Psi}+\bar{\Phi}$ \cite{Kaiser}.

\subsection{The energy-momentum tensor}
One can define a frame-invariant distribution function $F(x^i,P_j,\tau)$, giving the number of particles in a (frame-invariant) differential volume in phase space,
\beq
F(x^i,P_j,\tau) \,dx^1 dx^2 dx^3 dP_1 dP_2 dP_3 =d N\,.
\eeq
From $F$ one can define the comoving number density,
\beq
n_c(x^i,\tau) = g_s \int \frac{d^3P}{(2\pi)^3} \,F(x^i,P_j,\tau) \,,
\eeq
and the energy-momentum tensor,
\beq 
T_{\mu\nu} = g_s \int \frac{d^3P}{(2\pi)^3} (-g)^{-1/2}\,\frac{P_\mu P_\nu}{P^0} F(x^i,P_j,\tau) \,,
\eeq
where $d^3P = dP_1 dP_2 dP_3$ and $g_s$ counts the spin degrees of freedom. One can verify that the above definition fulfills the transformation rule of  eq.~(\ref{Ttrans}).
The distribution function for bosons (-) and fermions (+)  in equilibrium is the standard one \cite{MB},
\beq
F^0 (\epsilon) =   \left(e^{\epsilon/T_0}\pm 1 \right)^{-1}\,.
\eeq 

Using the variables $q_i$, $\epsilon$ defined in eq.~(\ref{qeps}), the components of the energy momentum tensor are explicitly given, at first order, by
\beqra
&&\bar{T}^0_0 = l_R^4\, T^0_0= - g_s \bar{a}^{-4} \int\frac{d^3q}{(2\pi)^3} \epsilon f = -  \bar{\rho} \left(1+\frac{\delta \bar{\rho}}{\bar{\rho}} \right)\,,\nonumber\\
&& \bar{T}^0_i=l_R^4\, T^0_i  =  g_s\bar{a}^{-4} \int\frac{d^3q}{(2\pi)^3} q n_i f =  \bar{\rho} (1+w) \, v^i\,,\nonumber\\
&&\bar{T}^i_j= l_R^4\, T^i_j =  g_s\bar{a}^{-4}  \int\frac{d^3q}{(2\pi)^3} \frac{q^2}{\epsilon}n_i n_j f
= \bar{\rho} \left[w +c_s^2\frac{\delta \bar{\rho}}{\bar{\rho}}\right] \delta^i_j+\Sigma^i_j\,,
\label{tmunucomp}
\eeqra
where $f(x^i,q,n_j,\tau) = F(x^i,P_j,\tau)$, $v^i\equiv d x^i/d\tau$, $w$ is the equation of state, and $c_s^2= \partial \bar{P}/\partial \bar{\rho}$.

\section{The Boltzmann equation}
\label{Boltzmann}
From what we have discussed in the previous session, it is now clear that, using the frame-invariant coordinates $x^i$, $P_j$, and $\tau$, it is possible to study departures from thermal equilibrium in a frame-invariant way. The tool is, as usual, the Boltzmann equation. The evolution of the phase space density of a particle $\psi$, $F_\psi(x_\psi^i,P^\psi_j,\tau)$ is given by
\beq
\frac{\partial F^\psi}{\partial \tau} +\frac{d x_\psi^i}{d \tau} \frac{\partial F^\psi}{\partial x_\psi^i} +\frac{d P^\psi_j}{d \tau} \frac{\partial F^\psi}{\partial P^\psi_j} = \left[\frac{d F^\psi}{d \tau} \right]_C\,.
\eeq
The frame-invariance of the LHS is trivially checked. One can cast it in a more useful form by using $dx_\psi^i/d\tau=P_\psi^i/P_\psi^0$ and eq.~(\ref{geod}). 

The collisional term for a generic process $\psi+a+b+\cdots\leftrightarrow i+j+\cdots$ reads
\beqra
&& \left[\frac{d F_\psi}{d \tau} \right]_C(x_\psi^i,P^\psi_j,\tau)\nonumber =\frac{1}{2P^\psi_0}
 \int d \Pi^a d \Pi^b\cdots d \Pi^i d \Pi^j\cdots \nonumber\\
 &&\times (2\pi)^4 \delta^4(P^\psi+P^a+P^b \cdots -P^i-P^j\cdots)\nonumber\\
&& \times\left[|{\cal M}|_{\psi+a+b+\cdots\rightarrow i+j+\cdots}^2   F_\psi F_a F_b\cdots (1\pm F_i) (1\pm F_j)\cdots \right.\nonumber\\
&& \left.- |{\cal M}|_{i+j+\cdots\rightarrow \psi+a+b+\cdots }^2 F_i F_j\cdots (1\pm F_\psi)(1\pm F_a)(1\pm F_b) \cdots \right]\,,
\label{collision}
 \eeqra
where the $''+''$ applies to bosons and the $''-''$ to fermions, and $d \Pi$ is the frame-invariant quantity
\beqra
&& d \Pi \equiv l_R^2 \,\frac{d^4P}{(2\pi)^3}\,(-g)^{-1/2} \, \delta(P^2+m^2) \Theta(P^0)\nonumber\\
&&\quad= l_R^2\, \frac{d^3P}{(2\pi)^3}\frac{(-g)^{-1/2}}{2P^0} =\frac{\bar{l}_R^2\,a^{-2}}{(2\pi)^3} \frac{d^3 q}{2 \epsilon}\,,
\eeqra
with $d^4P = dP_0 \, d^3P$.
The delta-function in eq.~(\ref{collision}) depends on momenta with low indices.

\subsection{Freeze out}
As a first example, we consider the case of a heavy particle decaying into two lighter ones, which are assumed to equilibrate rapidly. Following the standard procedure (see for instance ref.~\cite{Kolb}) the (conformal) time dependence of the comoving number density is given by
\beqra
&&\dot{n}_c^\psi = g_s \int\frac{d^3 P}{(2 \pi)^3}   \left[\frac{d F_\psi}{d \tau} \right]_C \nonumber \\
&&=\int \frac{d^3 P^\Psi}{(2 \pi)^3}  \frac{\left|{\cal M}\right|^2}{2P_0^\psi} d\Pi^a d\Pi^b (2\pi)^4 \delta^{(4)} (P^\psi-P^a-P^b) (F_\psi - F_a^0 F_b^0)\,,
\eeqra
where we have approximated $1 \pm F \simeq 1$.Using energy conservation we can write, as usual,
\beq
F_a^0 F_b^0=F_\psi^0\,,
\eeq
and then
\beq
\dot{n}_c^\psi = - (n_c^\psi-{n_c^\psi}^0) \int d\Pi^a d\Pi^b (2\pi)^4 \delta^{(4)} (P^\psi-P^a-P^b)   \frac{\left|{\cal M}\right|^2}{2 m_\psi a}\,,
\eeq
where for the non-relativistic particle $\Psi$ we have used $P_0^\psi\simeq -m_\psi a$.
Recognizing the integral as the decay rate per unit conformal time, $\Gamma\,a$, where $\Gamma$ is the decay probability per unit physical time, and turning to the variable $x=m_\psi/T$, we get the frame-invariant equation
\beq
\frac{d\, {n}_c^\psi}{d \log x} = -\frac{ \Gamma\,a}{H(1+m^\prime_\psi/m_\psi)} (n_c^\psi-{n_c^\psi}^0)\,,
\label{n1}
\eeq
where primes denote derivatives with respect to $\log a$ and we have used the relation $\dot{T}/T=-H$.
From the above equation one can see that the usual rule of thumb for a particle interaction to be efficient in the expanding Universe, that is $\Gamma\,a \agt H $, now generalizes to the frame-invariant relation 
\beq
\Gamma \,a \agt H(1+m^\prime_\psi/m_\psi)\,.
\eeq

We stress again that the frame independence of the above equation is a consequence of the frame-independence of the product $\Gamma a$, and then, ultimately, of that of the matrix element $\left|{\cal M}\right|^2$.

Analogously, in the case of a $2 \leftrightarrow 2$ scattering process one gets the result

\beq 
\frac{1}{{n_c^\psi}^0}\frac{d\, {n}_c^\psi}{d \log x} = -\frac{ \Gamma\,a}{H(1+m^\prime_\psi/m_\psi)} \left[\left(\frac{n_c^\psi}{{n_c^\psi}^0}\right)^2-1 \right]\,,
\label{n2}
\eeq
where $\Gamma = {n_c^\psi}^0 a^{-3} \langle \sigma v\rangle$, with $\langle \sigma v\rangle$ the thermally averaged cross section.

From these examples one can appreciate the utility of the frame-invariant Boltzmann equations. In practice, it turns out that rates, cross sections, and all the particle physics related quantities are more conveniently computed in a frame different from that in which the Einstein equations are simpler. For instance, in scalar-tensor theories, particle physics is conveniently computed in the Jordan frame, whereas gravitational equations are simpler in the Einstein frame. Since frame-invariant combinations --such as $\Gamma a$ -- appear in the equations above, 
one can first compute rates and cross sections in the more convenient frame and then translate them in the other one using the relations of eq.~(\ref{ncs}).

\subsection{Sachs-Wolfe effect}
Using the 0-0 component of the energy-momentum tensor, eq.~(\ref{tmunucomp}) one can define (space and direction-dependent) temperature fluctuations for a gas of photons ($\epsilon =q$, $g_s=2$) as
\beq
\Theta(x^i,n_j,\tau)  \equiv \frac{\Delta T}{T}(x^i,n_j,\tau) =  \frac{1}{4\pi^2\bar{\rho} \bar{a}^4} \int dq q^3 f -1\,.
\eeq
From the collisionless Boltzmann equation for the function $f$,
\beq
\frac{\partial f}{\partial \tau} +\dot{x}^i \,\frac{\partial f}{\partial x^i}+\dot{q}\,\frac{\partial f}{\partial q} + \dot{n}_j\, \frac{\partial f}{\partial n_j}=0\,,
\eeq
using the geodesic equation (\ref{geoq}) and the relation $n_i = q_i/q =\dot{x}^i (1-\bar{\Phi}-\bar{\Psi})\epsilon/q$, one gets 
\beq
\frac{d\;\;}{d\tau} (\Theta + \bar{\Psi}) = \dot{\bar{\Psi}}+\dot{\bar{\Phi}}\,,
\label{SW}
\eeq
where use has been made of the fact that potentials and $\delta l_R$ do not depend on the angle explicitly, and $\dot{(\,\,)}\equiv \partial\,/\partial\tau$.
If the potential are static, the quantity $\Theta +\bar {\Psi}$ is conserved, which is the frame-invariant expression for the Sachs-Wolfe effect \cite{SW}.

\subsection{Phase space evolution for CMB photons}
\label{cmb}
The evolution of the CMB photon distribution function is described by the Boltzmann equations discussed, for instance, in  \cite{BE,Ko,MB}. To reduce the number of variables, one integrates out the $q$ dependence and expands the angular dependence in Legendre  polynomials, $P_l$. Going to Fourier space, one defines
\beqra
F(\vec{k},\hat{n},\tau) &\equiv&   
\frac{\int f^1(\vec{k},\vec{q},\tau) q^3dq}{\int f^0(q)q^3dq} \equiv \nonumber\\
&\equiv& \sum^{\infty}_{l=0} (-i)^l (2l+1) F_{l}(\vec{k},\tau) P_{l}(\hat{k}\cdot\hat{n}) \,, \nonumber \\
G(\vec{k},\hat{n},\tau) &\equiv&   
\frac{\int Q^1(\vec{k},\vec{q},\tau) q^3dq}{\int Q^0(q)q^3dq} \equiv \nonumber\\
&\equiv& \sum^{\infty}_{l=0} (-i)^l (2l+1) G_{l}(\vec{k},\tau) P_{l}(\hat{k}\cdot\hat{n}) \,,\\
\label{variables}
\eeqra
where $\vec{k}=k \hat{k}$ is the wavevector and $\hat{n}$ the direction of the photons 3-momentum $\vec{q}$.
$f^0$ is the zeroth order (equilibrium) distribution function and $f^1$ the first order deviation from it, while $Q^0$ and $Q^1$ are the zeroth and first order Stokes parameter, respectively.

The Boltzmann equations for $F$ and $G$ take the form
\beqra
\frac{\partial F}{\partial\tau}+ i k\mu F-4(\dot{\bar{\Phi}}-i k \mu \bar{\Psi}) = \left( \frac{\partial F}{\partial \tau} \right)_C \,,\nonumber\\
\frac{\partial G}{\partial\tau}+ i k\mu G = \left( \frac{\partial G}{\partial \tau} \right)_C \,,
\eeqra
with $\mu\equiv \hat{k} \cdot \hat{n}$. Again, the LHS are manifestly frame-invariant.
The collisional terms are given by \cite{MB}
\beqra
\left( \frac{\partial F}{\partial \tau} \right)_C = a^{-2} n_c^e\, \sigma_T \left[ -F + F_{0} +4 \hat{n} \cdot \vec{v}_e
-\frac{1}{2} \left( F_2 + G_0 + G_2 \right) P_2 \right] \,, \nonumber\\ 
\left( \frac{\partial G}{\partial \tau} \right)_C = a^{-2} n_c^e \,\sigma_T \left[ -G 
+\frac{1}{2} \left( F_2 + G_0 + G_2 \right)(1-P_2)  \right] \,, 
\eeqra

where $\vec{v}_{e}$ and $n_c^e$ are respectively the proper velocity and comoving density of the electrons, and $\sigma_T$ the Thomson cross section.

Since $ a^{-2}n^e_c \, \sigma_T =\bar{a}^{-2} n^e_{c} \bar{\sigma}_T$ is frame-invariant, and so is the proper velocity, the frame-invariance of the collisional terms is also manifest.

\section{Equations of Motion for Scalar-Tensor theories}
\label{EoM}
It is convenient to define the frame-invariant metric
\beq
h_{\mu\nu} \equiv l_{Pl}^{-2} g_{\mu\nu}\,,
\label{hmunu}
\eeq 
where the unit length $l_{Pl}$ will be later identified with the Planck length.
If $g_{\mu\nu}$ is a FRW metric in Newtonian gauge, so is $h_{\mu\nu}$, with scale factor $\bar{a}$ and potentials $\bar{\Psi}$, $\bar{\Phi}$ as defined in eq.~(\ref{fisp}) with $l_R=l_{Pl}$. Notice that $h_{\mu\nu}$ and $\bar{a}^2$ have dimension of (mass)$^2$.

Scalar-tensor theories can be defined in terms of frame-independent quantities by the action
\beq
S=S_G[h_{\mu\nu},\,\varphi] + S_M[h_{\mu\nu} e^{-2 b[\vp]},\,\bar{\phi}\,,\bar{\psi}\,,\ldots;\,\bar{\lambda}_n]\,,
\label{action}
\eeq
where the frame-independent fields $\bar{\phi}\,,\bar{\psi}\,,\ldots$, and coupling constants $\bar{\lambda}_n$'s, appearing in the matter action $S_M$ are given by the combinations in eq.~(\ref{fi}) with $l_R=l_{Pl}$. 


The gravity action is given by
\beq
S_G= \kappa \int d^4 x \,\sqrt{-h}\left[ R(h) - 2 \,h^{\mu\nu} \partial_\mu\vp \,\partial_\nu\vp-4 \bar{V}(\vp)\right]\,.
\label{actiongrav}
\eeq

The only feature differentiating the action in eq.~(\ref{action}) from that of standard GR is the function $b[\vp(x)]$. In the $b=0$ limit, the scalar-tensor theory reduces to GR with an extra scalar field, $\vp$, which in this limit can be seen as an extra matter component minimally coupled to gravity. In this case, a constant $l_{Pl}$ can be univocally taken as the most convenient choice to measure all the dimensional quantities in the theory. In this units, both the Planck mass and particle masses, as well as atomic wavelengths, are constant. 

On the other hand, when $b\neq 0$, the scalar field $\vp(x)$ is non-minimally coupled to gravity and one has a genuine scalar-tensor theory. In the literature, these theories are usually discussed in two frames, the Einstein and the Jordan ones. In our language, choosing a frame corresponds to fixing the function $l_{Pl}(x)$ appropriately. 

The first possible choice is to take a constant $l_{Pl}$, which corresponds to the Einstein frame. The gravity action takes 
the usual Einstein-Hilbert form
\beq
S_G= \kappa\, l_{Pl}^{-2} \int d^4 x \,\sqrt{-g}\left[ R(g) - 2 \,g^{\mu\nu} \partial_\mu\vp \,\partial_\nu\vp-4 V(\vp)\right]\,,
\eeq
with $V=l_{Pl}^{-2} \,\bar{V}$. The combination in front of the integral fixes the Einstein-frame Planck mass, $\kappa\, l_{Pl}^{-2} = M_\ast^2/2 =(16 \pi G_\ast)^{-2}$. In other words, in this frame, dimensional units are set by the Planck scale. The matter action is obtained from the one of quantum field theory by substituting the Minkowsky metric $\eta_{\mu\nu}$ with $g_{\mu\nu} e^{-2b}$. Since in this frame the matter energy-momentum tensor is not conserved (see eq.~(\ref{tmnc})), particle physics quantities, like masses and wavelengths are not constant. 

The other choice corresponds to the Jordan frame, which is obtained by making the Planck length space-time dependent such as to reabsorb $b[\vp(x)]$ in $S_M$. This is accomplished if one choses $\tilde{l}_{Pl} = l_{Pl}\, e^{-b}$, where $l_{Pl}$ is the previously defined Planck length in the Einstein frame. With this choice the matter action takes the standard form of quantum field theory (with $\eta_{\mu\nu}\rightarrow g_{\mu\nu}$), whereas the gravity action is
\beq
S_G = \frac{M_\ast^2}{2} \int d^4 x \,\sqrt{-\tilde{g}} \,e^{2b} \left[ R(\tilde{g}) - 2 \,\tilde{g}^{\mu\nu} \partial_\mu\vp \,\partial_\nu\vp \,(1-3 \,\alpha^2)-4 \tilde{V}(\vp)\right]\,,
\eeq
where $\tilde{V}=\tilde{l}_{Pl}^{-2}\bar{V}$ and 
\beq
\alpha\equiv \frac{d b}{d\vp}\,.
\eeq
Notice that, in this frame, the r\^{o}le of the Planck mass is played by the space-time dependent quantity $M_\ast e^b$. Since $b[\vp(x)]$ disappears from the matter action, the energy-momentum tensor is now covariantly conserved.

Of course, any other choice for $l_{Pl}$ is possible in principle and leads to the same physical consequences. However, for practical purposes, only the Einstein and Jordan frames are employed. The discussion of the previous sections shows how one can exploit the good aspects of the two. One can compute cross sections, decay rates, etc., in the Jordan frame, where masses and couplings are constant and the usual rules of quantum field theory apply straightforwardly. Then, one constructs frame-invariant combinations out of these, such as $\tilde{\Gamma} \tilde{a}$,  and insert them into the frame-independent Boltzmann equations like eqs.~(\ref{n1}, \ref{n2}). On the other hand, the gravity part, like the combination $H(1+m^\prime/m)$ can be computed in the Einstein frame. Equivalently, both the particle physics part and the gravity part can be computed frame-invariantly from the beginning, using the particle physics parameters defined in eq.~(\ref{fi}) and solving the equations of motion 
obtained from the action in eq.~(\ref{action}), that we are going to write down explicitly up to first order.

Before doing that, we give the expression of the redshift (\ref{redshift_at}) in terms of the frame-invariant scale factor $\bar{a}=a/l_{Pl}$, that is
\beq
1+z=\frac{\bar{a}(\tau_0)}{\bar{a}(\tau)} e^{\bar{b}(\tau) -\bar{b}(\tau_0)}\,,
\label{rednew}
\eeq
where $\bar{b}(\tau)\equiv b[\bar{\vp}(\tau)]$.

\subsection{Background equations}
\label{bee}
The background equations for the scale factor $\bar{a}$ are
\beqra
&&\left(\frac{\dot{\bar{a}}}{\bar{a}}\right)^2-\frac{2}{3}\left(\frac{1}{2}\dot{\vp}^2 +\bar{a}^2 \bar{V}\right) = \frac{1}{6\kappa} \bar{\rho}\bar{a}^2 \,,\\
&&
\frac{\ddot{\bar{a}}}{\bar{a}} +\frac{1}{3}\left(\dot{\vp}^2 - 4\bar{a}^2 \bar{V}\right) =-\frac{1}{12\kappa} \bar{\rho}\bar{a}^2 (1-3w)\,.
\eeqra
The energy-momentum tensor $\bar{T}^\mu_\nu$ is not conserved,
\beq
\bar{T}^\mu_{\nu;\,\mu}=-\alpha \,\vp_\nu \,\bar{T}^\mu_{\mu}\,,
\label{tmnc}
\eeq
which, at zeroth-order implies
\beq
\dot{\bar{\rho}} +3\,\bar{\rho}\,(1+w)\, \dot{\bar{a}}/\bar{a} = -\alpha\, \dot{\vp} \,\bar{\rho}\, (1-3w)\,.
\eeq
One can verify that, with the Jordan frame choice, {\it i.e.} $\tilde{l}_{Pl}\sim e^{-b}$, the covariant conservation of the energy-momentum tensor is recovered.

The equation of motion for the field $\vp$ is
\beq
h^{\mu\nu} \bar{\cal D}_\nu\vp_\mu - \frac{\delta\bar{V}}{\delta \vp} = \frac{\alpha}{4 \kappa} \bar{T}^\mu_\mu\,,
\label{kg}
\eeq
which, at zeroth-order, gives 

\beq
\ddot{\varphi} +2\, \frac{\dot{\bar{a}}}{\bar{a}} \, \dot{\varphi} \,
+\bar{a}^2 \, \frac{\partial \bar{V}}{\partial \varphi} = \frac{\alpha}{4k} \, \bar{a}^2 \, \bar{\rho} \, (1-3w) \,. 
\eeq

\subsection{First-order equations}
\label{foe}
Here we give the set of first order equations:
\beqra
&& k^2 \Phib +3\frac{\dot{\ab}}{\ab}\left(\dot{\Phib}+ \frac{\dot{\ab}}{\ab} \Psib \right) - (\dot{\vp} \delta \dot{\vp}-\dot{\vp}^2 \Psib) +2 \ab^2\frac{\delta\bar{V}}{\delta \vp} \delta \vp=-\frac{1}{4\kappa} \delta \rhb \ab^2\,,\\
&& k^2 \left(\dot{\Phib}+ \frac{\dot{\ab}}{\ab} \Psib \right)-\dot{\vp} k^2 \delta\vp = \frac{1}{4\kappa}\rhb \ab^2 (1+w) \theta\,,\\
&& \ddot{\Phib} + \frac{\dot{\ab}}{\ab} (\dot{\Psib}+2\dot{\Phib}) + \left[2\frac{\ddot{\ab}}{\ab} - \left(\frac{\dot{\ab}}{\ab}\right)^2\right]\Psib -\frac{1}{3} k^2(\Psib-\Phib) +\nonumber\\
&&\qquad\qquad\qquad \qquad\qquad+ \Psib \dot{\vp}^2 -\dot{\vp}\delta\dot\vp + \frac{\delta\bar{V}}{\delta \vp} \delta \vp = \frac{1}{4\kappa} c_s^2 \delta \rhb\ab^2\,,\\
&& k^2(\Phib-\Psib) = \frac{3}{4 \kappa} \rhb\ab^2(1+w) \sigma\,,
\label{em4}
\eeqra
where $\theta\equiv i k^j v_j$ and $\rhb (1+w)\sigma\equiv - (\hat{k_i}\hat{k_j} - \delta_{ij}/3) \Sigma^i_j$, with $\hat{k_i}=k_i/k$.

At first order, the continuity  equation for the matter energy-momentum tensor, eq.~(\ref{tmnc}), yields 

\beqra
\dot{\bar{\delta}} &=& - (1+w)(\theta-3\dot{\bar{\Phi}})-(1-3w)\left( \alpha   \delta \dot{\varphi} 
+ \dot{\varphi} \frac{\partial\alpha}{\partial \varphi} \delta \varphi  \right) - \nonumber\\
&-& 3 \frac{\partial w}{\partial \bar{\rho}} \bar{\rho} \bar{\delta} \left( \frac{\dot{\bar{a}}}{\bar{a}} - \alpha \dot{\varphi}  \right) \,, \\
\dot{\theta} &=& -(1-3w)\left( \frac{\dot{\bar{a}}}{\bar{a}} - \alpha \dot{\varphi} \right) \theta - \frac{\dot{w}}{1+w} \theta
+\frac{\left(w+\frac{\partial w}{\partial \bar{\rho}} \bar{\rho} \right)  }{1+w} k^2 \bar{\delta} \nonumber\\
&+& k^2 (\bar{\psi}-\sigma) - \frac{(1-3w)}{(1+w)} k^2 \alpha \delta \varphi \,,
\eeqra

where $\bar{\delta}\equiv \delta \bar{\rho}/\bar{\rho}$. \\
The equation of motion for the scalar field fluctuation $\delta \vp$, from eq.~(\ref{kg}) is

\beqra
\delta \ddot{\varphi} &+& 2\frac{\dot{\bar{a}}}{\bar{a}} \delta \dot{\varphi}
-\nabla^2 \delta \varphi  -\dot{\bar{\psi}} \dot{\varphi}
-3\dot{\varphi}\dot{\bar{\Phi}} +2 \bar{\psi} \bar{a}^2 \frac{\partial \bar{V}}{\partial \varphi} 
-\frac{\alpha}{2k} \bar{\rho}\bar{a}^2 (1-3w)\bar{\psi} + \nonumber\\
&+& \bar{a}^2 \frac{\partial^2 \bar{V}}{\partial \varphi^2} \delta \varphi = 
 \frac{\bar{a}^2}{4k} \left[ \bar{\rho} (1-3w) \frac{\partial \alpha}{\partial \varphi} \delta \varphi +
\left( 1-3w -3\frac{\partial w}{\partial \bar{\rho}} \bar{\rho}  \right) \alpha \delta \bar{\rho}  \right] \,.
\eeqra

\section{Conclusion}
A particular ST theory is identified by two functions: $b(\vp)$ and the effective potential $\bar{V}(\vp)$, see eqs.~(\ref{action}) and (\ref{actiongrav}). Therefore, the physical deviations from GR should be parameterized in terms of these two functions alone, irrespectively of the frame one choses to solve the equations of motion \cite{pn,dam}. Actually, as we have shown, fixing a frame is not necessary, provided one carefully expresses all the observables in terms of frame-invariant quantities.

From a practical point of view, the formulation of ST gravity presented in this paper allows a straightforward modification of the available codes based on Boltzmann equations for the study of Nucleosynthesis, CMB, or the calculation of the Dark Matter relic abundances in the context of GR. It is enough to redefine the redshift as in eq.~(\ref{rednew}) and add the scalar field $\vp$ to the GR equations of motion, as in sects.~\ref{bee} and \ref{foe}. Then, the code will work in the standard way, the only difference being given by the two extra inputs $b$ and $\bar{V}$. The implementation of this procedure to the publicly available CMBFAST \cite{cmbfast} code is under way.

\section{Appendix}
We will show how the results obtained in the text in the Newtonian gauge can be extended  to the synchronous gauge and to a generic gauge.

\subsection{The synchronous gauge}

The synchronous gauge is defined by

\beq
ds^2=a^2 \left[ -d\tau^2 + \left(\delta_{ij} + h_{ij} \right)dx^i dx^j  \right] \,.
\label{sg1}
\eeq

After a frame transformation $ds^2 \to d\tilde{s}^2=e^{-2f}ds^2$ 
the metric of eq.~(\ref{sg1}) transforms in

\beq
d\tilde{s}^2= a^2 e^{-2f_{B}} \left[ -(1-2\delta f)d\tau^2 + \left(\delta_{ij} + h_{ij} - 2\delta f \delta_{ij} \right)dx^i dx^j  \right] \,,
\eeq
Unlike for the Newtonian gauge, a frame transformation doesn't preserve the synchronous gauge.
However it is possible to give a frame-independent description of all the 
physical phenomena also when the metric perturbations are described in the basis $(h_{ij},\delta f)$. 
In fact, as we will see, the geodesic motion depends only from the frame-independent combination
$\bar{h}_{ij}= h_{ij} - 2 (\delta l_R/l_R) \delta_{ij}$.\\
To show this in a simple way let's define a synchronous gauge as in the following

\beq
dh^2= \bar{a}^2 \left[ -d\tau^2 + \left(\delta_{ij} + \bar{h}_{ij} \right)dx ^i dx^j  \right] \,.
\label{sg2}
\eeq

In this way at least the spatial components of the metric are frame-independent.\\
As for the Newtonian gauge we now relate the 4-momentum $P_{\mu}$ to the frame-independent variables $q^i$ and $\epsilon$

\beqra
P_{0}&=&- \epsilon \,, \nonumber\\
P_{i}&=&\left( \delta_{ij} + \frac{1}{2} \bar{h}_{ij}   \right)q^j \,.
\label{mo}
\eeqra

With this definitions the relation $\epsilon= \sqrt{q^2+\bar{m} \bar{a}^2}$ is preserved.\\
Using eqs.~(\ref{mo}) and the metric of eq.~(\ref{sg2}) the geodesics equation at first order yields

\beq
\dot{q} =  - \frac{1}{2} q n^i n^j \dot{\bar{h}}_{ij} \,.
\label{geosy}
\eeq

Also if the metric of eq.~(\ref{sg2}) is not frame-invariant, eq.~(\ref{geosy}) is not affected by a frame transformation. 

\subsection{The case of a generic gauge}

Let's consider now the metric

\beq
ds^2=a^2 \{ -(1+2 \psi)d\tau^2 +2\partial_i B d\tau dx^i +\left[ (1-2\Phi)\delta_{ij} + D_{ij}E \right]dx^i dx^j  \}
\label{gega}
\eeq

where $D_{ij}=\left(\partial_i \partial_j -\frac{1}{3} \delta_{ij} \nabla^2\right)$.\\
After a frame transformation $ ds^2 \to  d\tilde{s}^2=e^{-2f} ds^2$ eq.~(\ref{gega}) transforms in

\begin{eqnarray}
\lefteqn{
d\tilde{s}^2 = a^2 e^{-2f_{B}} \{ -(1+2 \psi-2\delta f)d\tau^2 +2\partial_i B d\tau dx^i +{} }  \nonumber\\ 
& & {} + \left[ (1-2\Phi-2\delta f)\delta_{ij} + D_{ij}E \right]dx^i dx^j  \} 
\end{eqnarray}

The transformation properties of the metric in eq.~(\ref{gega}) suggest to define the
following frame-invariant line element 

\beq
dh^2=\bar{a}^2 \{ -(1+2 \bar{\psi})d\tau^2 +2\partial_i \bar{B} d\tau dx^i +\left[ (1-2\bar{\Phi})\delta_{ij} + D_{ij} \bar{E} \right]dx^i dx^j  \} \,,
\label{gegafi}
\eeq

where the frame-invariant quantities $\bar{a}$, $\bar{\psi}$ and $\bar{\Phi}$ are given in eq.~(\ref{fisp}). 
We also wrote $\bar{B}=B$ and $\bar{E}=E$ to underline that such a quantities do not transform under frame transformations.\\
The 4-momentum is now related to the frame invariant variables $q^i$ and $\epsilon$ by the following relations

\beqra
P_{0}&=&-\left[q^i \partial_i \bar{B} + \epsilon(1+\bar{\psi}) \right] \,, \nonumber\\
P_{i}&=&\left[ (1-\bar{\Phi})\delta_{ij} + \frac{1}{2} D_{ij} \bar{E}  \right]q^j \,,
\label{momenta}
\eeqra

with $\epsilon= \sqrt{q^2+\bar{m} \bar{a}^2}$.\\
Using now eqs.~(\ref{momenta}) and the metric of  eq.~(\ref{gegafi}) at first order the geodesic equation yields 

\beqra
\dot{q} = q\dot{\bar{\Phi}} - \epsilon n^i \partial_i \bar{\psi} + 2\epsilon n^i \left( \frac{\dot{\bar{a}}}{\bar{a}} \partial_i \bar{B}
+ \partial_i \dot{\bar{B}}  \right) - q n^i n^j \left( \partial_i \partial_j \bar{B} + \frac{1}{2} D_{ij}\dot{\bar{E}} \right)
\label{geoge}
\eeqra

If in eq.~(\ref{geoge}) we impose $\bar{B}=\bar{E}=0$ we recover eq.~(\ref{geod}).
Choosing instead $\bar{B}=\bar{\psi}=0$ and $-2\bar{\Phi}\delta_{ij}+D_{ij}\bar{E}=\bar{h}_{ij}$  
eq.~(\ref{geoge}) reduces to  eq.~(\ref{geosy}).

\acknowledgments 

R. Catena acknowledges a Research Grant funded by the VIPAC Institute.

\end{document}